\documentclass[18pt,a4paper,final,twoside]{report}
\usepackage[utf8x]{inputenc}
\usepackage{ucs}
\usepackage[english]{babel}
\usepackage{amsmath}
\usepackage{amsfonts}
\usepackage{amssymb}
\usepackage{mathrsfs}
\usepackage{makeidx}
\usepackage{graphicx}
\usepackage{lmodern}

\newcommand{\trace}{\text{tr}}
\newcommand{\Trace}{\text{Tr}}
\newcommand{\Leff}{\mathscr{L}}
\newcommand{\F}{\mathcal{F}}
\newcommand{\G}{\mathcal{G}}

\title{Some remarks on the use of effective Lagrangians in QED and QCD}

\begin{document}

\begin{titlepage}

\begin{center}
{\Large Some remarks on the use of effective Lagrangians in QED and QCD}\\
\vspace{5em}

\large
Walter Dittrich\\
\makeatletter
Institute for Theoretical Physics\\
University of T\"{u}bingen\\
Auf der Morgenstelle 14\\
D-72076 T\"{u}bingen\\
Germany\\
qed.dittrich@uni-tuebingen.de\\
\makeatother
\end{center}
\end{titlepage}

\begin{abstract}
We discuss in this article the usefulness of the effective Lagrangians ($\mathscr{L}^{eff}$) of QED
and QCD within the one-loop approximation. Instead of calculating $\mathscr{L}^{eff}$ via complicated
computations with Schwinger's proper-time technique or Feynman graphs,
we prefer to employ the energy-momentum tensor and the leading-log model. The advantage is
that we do not have to demand the external electromagnetic or color field to be constant.
There are also some critical remarks added which cast doubt on the use of $\mathscr{L}_{QCD}$ with
covariant constant fields in explaining the nature of the QCD vacuum.
\end{abstract}

\section{Introduction}
In the first chapter we compile the most important results with regard to the effective Lagrangian in a constant electromagnetic field.
Our objective is to find the Green's function of a spin-$\tfrac{1}{2}$ particle in an external constant magnetic field
that points in the $\hat{z}$ direction. This can be achieved with Schwinger's proper-time technique.
With the result we can compute the entire effective Lagrangian as a function of the constant $(E,H)$ field. In this way we obtain the
famous Heisenberg-Euler effective Lagrangian. We will then set up a relation between the effective Lagrangian and the trace
of the energy-momentum tensor for constant magnetic and electric fields. Thereafter we give up the requirement that the fields be constant
and allow for arbitrary varying fields. This is done in connection with the effective action for Yang-Mills fields.
Rather than attempting to compute $\Leff^{eff}$, we will make an ansatz motivated by the requirement that $\Leff^{eff}$
give the correct trace anomaly for the energy-momentum tensor. In this way we are able to construct the leading-log effective Lagrangian.
Similar considerations are used to investigate the effective Lagrangian in QED. Finally we briefly study Adler's leading-log model in QCD
and state his result concerning the static potential between a quark-antiquark pair for long and short distances. Although the calculations are
highly non-trivial, the results of the linearly rising potential for large quark separation and the Coulombic $r^{-1}$ potential for small
distances are very encouraging.

\section{Compendium of Useful Formulae\cite{Dittrich1985} }
We start with the Green's function of a spin-$\tfrac{1}{2}$ particle in an external electromagnetic field:
\begin{equation}
\left[ \gamma^\mu\left(\frac{1}{i}\partial_\mu - eA_\mu\right) + m\right]G_+(x,x';A) = \delta(x-x').
\end{equation}
If we pick a special gauge field so that $F_{\mu\nu}$ is constant, we obtain the closed-form solution
\begin{align*}
G_+(x,x';A) = \phi(x,x') &\int_0^\infty \frac{1}{s^2}\left[m-\frac{1}{2}\gamma^\mu\left\lbrace f(s)+ eF\right\rbrace_{\mu\nu}(x-x')^\nu\right]\\
&\times e^{-im^2s-L(s)+\frac{i}{4}(x-x')f(s)(x-x')}e^{\frac{i}{2}\sigma Fs}ds\quad ,
\intertext{with}
f(s) &= eF\coth(eFs)\\
L(s) &= \frac{1}{2}\trace \ln \left[ \frac{\sinh(eFs)}{eFs} \right]\ ,
\end{align*}
and $\phi(x,x') = e^{ie\int_{x'}^x A^\mu(\xi)d\xi_\mu }$ with a straight path between $x$ and $x'$.

Our central subject of interest is the vacuum amplitude in the presence of an external field which,
in the framework of a one-loop approximation for the effective Lagrangian, can be written as
\begin{align}
\langle 0_+\vert0_-\rangle^A &= e^{iW^{(1)}\left[A\right]} = e^{i\int\mathscr{L}^{(1)}(x)d^4x}\quad ,\\
\intertext{with}
iW^{(1)}\left[A\right] &= -\Trace\ln\left(\frac{1}{1-e\gamma AG_+}\right)=-\Trace\ln\left(\frac{G_+\left[A\right]}{G_+\left[0\right]}\right)\ .
\end{align}
Here $G_+=G_+[0]$ is the electron propagator in the field-free case, connected with $G_+[A]$ by
\begin{equation}
G_+[A] = G_+(1-e\gamma AG_+)^{-1}\quad .
\end{equation}
Furthermore, $\Trace$ indicates the trace both in spinor and configuration space.

The one-loop effective action $W^{(1)}$, i.e., the effective Lagrangian $\mathscr{L}^{(1)}$, is the formal expression for the effect which an arbitrary number of 
\textquotedblleft external photon lines\textquotedblright\ can have on a single Fermion loop.

The functional derivative with respect to the potential $A_\mu(x)$ is given by
\begin{equation}
i\frac{\delta W^{(1)}[A]}{\delta A_\mu(x)} = -e\trace\left[\gamma^\mu G_+(x,x;A)\right]\quad .
\end{equation}
This equation is fulfilled by the ansatz
\begin{equation}
iW^{(1)} := i\int\mathscr{L}^{(1)} d^4x = -\frac{1}{2}\int_0^\infty \frac{e^{-ism^2}}{s}\Trace\left[e^{is(\gamma\cdot \Pi)^2}\right]ds\ ,
\end{equation}
where the proper-time representation of $G_+[A]$ is given by
\begin{equation}
G_+[A]\cdot \frac{\gamma\cdot \Pi - m}{(\gamma\cdot \Pi)^2 - m^2} = (m-\gamma\cdot \Pi)i\int_0^\infty e^{-is\left[m^2-(\gamma\cdot\Pi)^2\right]}ds\quad .
\end{equation}
We can then write for the unrenormalized Lagrangian
\begin{equation}
\mathscr{L}^{(1)}(x) = \frac{i}{2}\trace \int_0^\infty \frac{e^{-im^2s}}{s}\langle x\vert e^{is(\gamma\cdot\Pi)^2} \vert x\rangle ds\ ,
\end{equation}
where the trace refers only to the spinor index.
With this expression for $\mathscr{L}^{(1)}(x)$ we can show that
\begin{equation}
i\frac{\partial \mathscr{L}^{(1)} }{\partial m}= \trace\ G_+(x,x;A)\ .
\end{equation}
Without further proof we also find for the trace of the energy-momentum tensor
\begin{equation}
\langle T_\mu ^\mu(x)\rangle = -im\ \trace\ G_+(x,x;A)\ .
\end{equation}
This leads us to the equation
\begin{align*}
\langle T_\mu ^\mu(x)\rangle &= -im\trace\langle x\vert (-\gamma\cdot\Pi+m)i\int_0^\infty e^{-i\left(m^2-(\gamma\cdot\Pi)^2\right)s}ds \vert x\rangle\\
&= m^2 \langle x\vert \trace \int_0^\infty e^{-i\left(m^2-(\gamma\cdot\Pi)^2\right)s} ds\vert x\rangle\ .
\end{align*}
With the former expression for $\mathscr{L}^{(1)}$ we obtain the useful equality
\begin{equation}
\langle T_\mu ^\mu(x)\rangle = m\frac{\partial \mathscr{L}^{(1)} (x)}{\partial m} = \frac{\partial \mathscr{L}^{(1)} (x) }{\partial(\ln m)}\quad .
\end{equation}
For a purely constant magnetic field the renormalized one-loop effective Lagrangian is known to be
\begin{equation}
\mathscr{L}^{(1)}(H) = -\frac{1}{8\pi^2}\int_0^\infty \frac{e^{-m^2s}}{s^2}\left[ (eHs)\coth(eHs)-\frac{1}{3}(eHs)^2-1\right]ds\ .
\end{equation}
The integral can be explicitly calculated by dimensional or $\zeta$-function regularization. In the next chapter we will make the explicit expression
for $\mathscr{L}^{(1)}(H)$ the starting point for our detailed discussion of the trace anomaly of the energy-momentum tensor in QED.

\section{The trace anomaly of the energy-momentum tensor from the one-loop effective Lagrangian in QED}
We already mentioned the close connection between the effective Lagrangian $\mathscr{L}^{(1)}$ and the trace of the energy-momentum tensor:
\begin{equation}
\langle T_\mu ^\mu(x)\rangle = m\frac{\partial \mathscr{L}^{(1)} (x) }{\partial m}\ .
\end{equation}
For constant fields we have the expression
\begin{align*}
\langle T_\mu ^\mu(x)\rangle \left(\mathcal{F},\mathcal{G}\right) &=  \frac{m^2}{16\pi^2}4\int_0^\infty \frac{e^{-m^2s}}{s^2}\left[ e^2s^2\mathcal{G}
\frac{ \mathfrak{Re}\cosh\left(es\sqrt{2}(\mathcal{F}+i\mathcal{G})^{1/2}\right) }{ \mathfrak{Im}\cosh\left(es\sqrt{2}(\mathcal{F}+i\mathcal{G})^{1/2}\right) } \right.\\
&\left.-1 -\frac{2}{3}e^2s^2\mathcal{F}\right]ds\quad ,
\intertext{where}
\mathcal{F} &= \frac{1}{4}F_{\mu\nu}F^{\mu\nu} = \frac{1}{2}\left( H^2-E^2\right)\ ,\\
\mathcal{G} &= \frac{1}{4}F^{\mu\nu}F_{\mu\nu} = \vec{E}\cdot \vec{H} \quad ,
\intertext{with}
F_{\mu\nu} &= \frac{1}{2}\epsilon_{\mu\nu\kappa\lambda}F^{\kappa\lambda},\quad \epsilon_{0123}=1\ .
\end{align*}

The closed-form expression $\mathscr{L}^{(1)}$ for an external constant $H$-field only is given by
\begin{align*}
\mathscr{L}^{(1)}(H) &= -\frac{1}{32\pi^2}\left[(2m^4-4m^2(eH)+\frac{4}{3}(eH)^2)\left[1+\ln\left(\frac{m^2}{2eH}\right)\right] \right.\\
&\left. +4m^2(eH)-3m^4-(4eH)^2\zeta'\left(-1,\frac{m^2}{2eH}\right)\right]\quad .
\end{align*}
The result of the mass-differentiation turns out to be
\begin{align*}
\langle T_\mu^\mu\rangle(H) &= -\frac{1}{12\pi^2}(eH)^2-\frac{m^4}{4\pi^2}\ln \left( \frac{m^2}{2eH} \right)+\frac{m^2}{4\pi^2}(eH)\ln\left( \frac{m^2}{2eH} \right)\\
& + \frac{m^4}{4\pi^2}+\frac{(eH)m^2}{2\pi^2}\left[\ln\Gamma\left(\frac{m^2}{2eH}\right)-\frac{1}{2}\ln2\pi\right]\quad .
\end{align*}
This, by the way, is also the result of the calculation of the integral
\begin{equation}
\langle T_\mu^\mu\rangle(H) = -i\frac{eHm^2}{4\pi^2}\int_0^\infty \frac{e^{-2ihz} }{z^2}\left[z\cot z-1+\frac{1}{3}z^2\right]dz,\quad h=\frac{m^2}{2eH}\ .
\end{equation}
Now observe that for $h\ll 1$ we can approximate $\ln\Gamma(h)\approx -\ln h$, such that
\begin{equation}
\lim_{m\rightarrow 0} \langle T_\mu^\mu\rangle(H) = -\frac{1}{12\pi^2}e^2H^2\ ,
\end{equation}
which, when written covariantly, yields
\begin{equation}
m\rightarrow 0\ :\quad \langle T_\mu^\mu\rangle = -\frac{1}{24\pi^2}e^2F_{\mu\nu}F^{\mu\nu} = -\frac{2\alpha}{3\pi}\frac{1}{4}F_{\mu\nu}F^{\mu\nu}\ .
\end{equation}
We can also obtain the next-to-leading term,
\begin{equation}
\langle T_\mu^\mu\rangle = -\beta(\alpha)\frac{1}{4}F_{\mu\nu}F^{\mu\nu},\quad \beta(\alpha)=\frac{2}{3}\left(\frac{\alpha}{\pi}\right)+\frac{1}{2}\left(\frac{\alpha}{\pi}\right)^2\ ,
\end{equation}
by incorporating results for the two-loop calculation $\mathscr{L}^{(2)}$.

For large field strengths, $\frac{eH}{m^2}\gg 1$, the dominant term is
\begin{equation}
\langle T_\mu^\mu\rangle(H) = -\frac{\alpha}{3\pi}H^2\ ,
\end{equation}
while for small field strength, $\frac{eH}{m^2}\ll 1$, we obtain, using Stirling's (Moivre's) formula for the logarithm of the $\Gamma$-function,
\begin{equation}
\langle T_\mu^\mu\rangle(H)=4\left(-\frac{2\alpha^2}{45}\frac{H^4}{m^4}+\frac{64}{315}\pi\alpha^3\frac{H^6}{m^8}+\dots\right)\ .
\end{equation}
Interestingly, the first term in this expansion agrees with Schwinger's\cite{Schwinger1951} from the Heisenberg-Euler Lagrangian:
\begin{equation}
T_{\mu\nu} = T_{\mu\nu}^{Maxwell}\left(1-\frac{16}{45}\alpha^2\frac{1}{m^4}\mathcal{F}\right) - \delta_{\mu\nu}\frac{2}{45}\alpha^2\frac{1}{m^4}\left(4\mathcal{F}^2
+\mathcal{F}\mathcal{G}^2\right)\ .
\end{equation}
In our present case we use $\hbar=c=1$, and for $\vec{E}=\vec{0}$ we have $\mathcal{F}=\tfrac{1}{2}(H^2-E^2)=\tfrac{1}{2}H^2$ and $\mathcal{G}=\vec{E}\cdot \vec{H}=0$.

Since we are interested in the trace of $T_{\mu\nu}$, we obtain (in Schwinger's notation)
\begin{align*}
\langle T_{\mu\mu}\rangle &= -4\left(\frac{8\alpha^2}{45m^4}\mathcal{F}^2+\frac{14\alpha^2}{45m^4}\mathcal{G}^2\right)\ ,
\intertext{which for $\mathcal{G} = 0$ indeed yields}
\langle T_{\mu\mu}\rangle(H) &= -4\left(\frac{2\alpha^2}{45}\frac{H^4}{m^4}\right)\ .
\end{align*}

Let us prove Schwinger's formula. He starts with
\begin{equation}
T_{\mu\nu} = \delta_{\mu\nu}\mathscr{L} - 2\frac{\partial\mathscr{L}}{\partial F_{\mu\lambda}}F_{\nu\lambda}\ .
\end{equation}
(Note that in Schwinger's formula the factor $2$ is missing!)

We need the following derivatives:
\begin{align*}
\frac{\partial\Leff (\F,\G) }{\partial F_{\mu\lambda}} &= \frac{\partial\Leff}{\partial\F}\frac{\partial\F}{\partial F_{\mu\lambda}}
+\frac{\partial\Leff}{\partial\G}\frac{\partial\G}{\partial F_{\mu\lambda}},\\
\frac{\partial\F}{\partial F_{\mu\lambda}} &=\frac{\partial}{\partial  F_{\mu\lambda}}\left(\frac{1}{4}F^2_{\rho\sigma}\right)=\frac{1}{2} F_{\mu\lambda}\\
\frac{\partial\G}{\partial  F_{\mu\lambda}} &= \frac{\partial}{\partial  F_{\mu\lambda}}\left(\frac{1}{4}F_{\rho\sigma}F^{*}_{\rho\sigma}\right) 
= \frac{1}{4}\frac{\partial}{\partial F_{\mu\lambda}}\left( F_{\rho\sigma}\frac{i}{2}\epsilon_{\rho\sigma\tau\omega}F_{\tau\omega}\right) \\
&= \frac{1}{2}  F_{\lambda\mu}^{*}\quad .
\end{align*}
Hence we can write
\begin{align*}
\frac{\partial\Leff (\F,\G) }{\partial  F_{\mu\lambda}} &= \frac{\partial\Leff}{\partial\F}\frac{1}{2}F_{\mu\lambda} + \frac{\partial\Leff}{\partial\G}\frac{1}{2}F_{\lambda\mu}^{*}\ ,
\intertext{so that}
-2\frac{\partial\Leff}{\partial F_{\mu\lambda}}F_{\nu\lambda} &= -\frac{\partial\Leff}{\partial\F}F_{\mu\lambda}F_{\nu\lambda} - \frac{\partial\Leff}{\partial\G} \underbrace{F_{\lambda\mu}^{*}F_{\nu\lambda}}_{=\G\delta_{\mu\nu}}\\
&= -F_{\mu\lambda}F_{\nu\lambda}\frac{\partial\Leff}{\partial\F} - \delta_{\mu\nu}\G\frac{\partial\Leff}{\partial\G}+\left(\delta_{\mu\nu}\F\frac{\partial\Leff}{\partial\F}
-\delta_{\mu\nu}\F\frac{\partial\Leff}{\partial\F}\right)\ .
\end{align*}
Putting everything together we obtain
\begin{align*}
T_{\mu\nu} &= -F_{\mu\lambda}F_{\nu\lambda}\frac{\partial\Leff}{\partial\F}+\delta_{\mu\nu}\F\frac{\partial\Leff}{\partial\F} + \delta_{\mu\lambda}\Leff 
- \delta_{\mu\nu}\F\frac{\partial\Leff}{\partial\F} - \delta_{\mu\nu}\G\frac{\partial\Leff}{\partial\G}\\
&= - \underbrace{\left( F_{\mu\lambda}F_{\nu\lambda}-\delta_{\mu\nu}\frac{1}{4}F_{\lambda\kappa}^2\right)}_{=T_{\mu\nu}^M}\frac{\partial\Leff}{\partial\F}
+\delta_{\mu\nu}\left(\Leff - \F\frac{\partial\Leff}{\partial\F}-\G\frac{\partial\Leff}{\partial\G}\right)\ ,
\end{align*}
which is a gauge-invariant expression.

Now, from the Heisenberg-Euler effective Lagrangian we are given
\begin{equation}
\Leff = -\F+C\left[4\F^2+7\G^2\right],\ C=\frac{2\alpha^2}{45m^4},\ \hbar=c=1\ .
\end{equation}
From this expression we obtain the derivatives
\begin{align*}
\frac{\partial\Leff}{\partial\F} &= -1+8C\F\quad ,& \F\frac{\partial\Leff}{\partial\F} &= -\F+8C\F^2\\
\frac{\partial\Leff}{\partial\G} &= 14C\G\quad ,& \G\frac{\partial\Leff}{\partial\G} &= 14C\G^2\ .
\end{align*}
Finally we end up with
\begin{align*}
T_{\mu\nu} &= T_{\mu\nu}^M \left( 1- \frac{16\alpha^2}{45m^4}\F\right)+\delta_{\mu\nu}\left(\Leff+\F-8C\F^2-14C\G^2\right),\left( \Leff = -\F +C(4\F^2+7\G^2)\right)\\
&= T_{\mu\nu}^M\left(1-\frac{16\alpha^2}{45m^4}\F\right) -\delta_{\mu\nu}\frac{2\alpha^2}{45m^4}\left(4\F^2+7\G^2\right)\qquad \square
\end{align*}
Let us put things together. Besides $\langle T_\mu^\mu\rangle(H)$, we can easily produce the corresponding result for a constant electric field by substituting
$H\rightarrow -iE$. The result is
\begin{align*}
\langle T_\mu^\mu\rangle(E) &= \frac{e^2E^2}{12\pi^2}-\frac{m^4}{4\pi^2}\left(i\frac{\pi}{2}+\ln\frac{m^2}{2eE}\right) -i\frac{eEm^2}{4\pi^2}\left(i\frac{\pi}{2}+\ln\frac{m^2}{2eE}\right)\\
&+\frac{m^4}{4\pi^2} -i \frac{eEm^2}{2\pi^2}\left[\ln\Gamma\left(\frac{im^2}{2eE}\right) - \frac{1}{2}\ln2\pi\right]\quad .
\end{align*}
If we split this equation up into its real and imaginary part we obtain
\begin{align*}
\mathfrak{Re}\langle T_\mu^\mu\rangle(E) &= \frac{e^2E^2}{12\pi^2}-\frac{m^4}{4\pi^2}\ln\frac{m^2}{2eE}+\frac{m^2eE}{8\pi}+\frac{m^4}{4\pi^2}
+\frac{eEm^2}{2\pi^2}\mathfrak{Im}\ln\Gamma\left(\frac{im^2}{2eE}\right)\\
\mathfrak{Im}\langle T_\mu^\mu\rangle(E) &= -\frac{m^4}{8\pi}-\frac{m^2}{4\pi^2}eE\ln\frac{m^2}{2eE}-\frac{eEm^2}{2\pi^2}\left[\mathfrak{Re}\ln\Gamma\left(\frac{im^2}{2eE}\right)
-\frac{1}{2}\ln 2\pi\right]\ .
\end{align*}
The last expression can be simplified with the aid of
\begin{equation}
\mathfrak{Re}\ln\Gamma(i\alpha) =\ln\vert\Gamma(i\alpha)\vert = \frac{1}{2}\ln\vert\Gamma(i\alpha)\vert^2 = -\frac{1}{2}\ln\left( \frac{\alpha\sinh(\pi\alpha)}{\pi}\right)\ .
\end{equation}
The result is
\begin{equation}
\mathfrak{Im}\langle T_\mu^\mu\rangle(E) = -\frac{m^4}{8\pi}+\frac{eEm^2}{4\pi^2}\ln\left[2\sinh\frac{\pi m^2}{2eE}\right]\ .
\end{equation}

Let us study these expressions in the limiting case $m\rightarrow 0$. To do this we employ the asymptotic formulae (for $z\ll 1$):
\begin{align*}
\ln\Gamma(z) &\approx \ln z\\
\mathfrak{Im}\ln\Gamma(iz) &\approx Cz,\quad \left( C\approx 0.577216\right)\\
\ln\sinh z &\approx \ln z\quad .
\intertext{We then obtain}
\lim_{m\rightarrow 0} \langle T_\mu^\mu\rangle(H) &= -\frac{1}{12\pi^2}e^2H^2\\
\lim_{m\rightarrow 0} \mathfrak{Re}\langle T_\mu^\mu\rangle(E) &= \frac{1}{12\pi^2}e^2E^2\\
\lim_{m\rightarrow 0} \mathfrak{Im}\langle T_\mu^\mu\rangle(E) &= 0\ .
\end{align*}
These three results are contained in
\begin{equation}
\lim_{m\rightarrow 0} \langle T_\mu^\mu\rangle = -\frac{1}{24\pi^2}e^2F_{\mu\nu}F^{\mu\nu}\ .
\end{equation}
We thus obtain a confirmation of the more general formula (in one-loop approximation)
\begin{equation}
\langle T_\mu^\mu (x) \rangle = -m\langle \bar{\psi}(x)\psi(x)\rangle -\frac{1}{24\pi^2}e^2F_{\mu\nu}(x)F^{\mu\nu}(x)\ ,
\end{equation}
where $\lim_{m\rightarrow 0} \left(m\langle \bar{\psi}(x)\psi(x)\rangle\right) = 0$.

Let us have a final look at $\mathfrak{Im} \langle T_\mu^\mu\rangle(E)$, and write it in units of $E^2_{cr} :=\tfrac{m^4}{4\pi\alpha}$, and the electric field 
in units of $E_{cr} := \frac{m^2}{e}$. Thus we obtain
\begin{align*}
\mathfrak{Im} \langle T_\mu^\mu\rangle(E) &= -\frac{\alpha}{2}+\frac{\alpha}{\pi}E\ln\left[2\sinh\frac{\pi}{2E}\right]\\
&= -\frac{\alpha}{2}+\frac{\alpha}{\pi}E\ln\left[e^{\frac{\pi}{2E}}-e^{-\frac{\pi}{2E}}\right]\\
&= -\frac{\alpha}{2}+\frac{\alpha}{\pi}E\left[\ln e^{\frac{\pi}{2E}}+\ln\left(1-e^{-\frac{\pi}{E}}\right) \right]\ .
\end{align*}
Here we use $\ln(1-x)=-\sum_{n=1}^\infty \frac{x^n}{n},\ -1\leq x < 1$, which yields
\begin{equation}
\mathfrak{Im} \langle T_\mu^\mu\rangle(E) = -\frac{\alpha}{2}+\frac{\alpha}{2}-\frac{\alpha}{\pi}E\sum_{n=1}^\infty\frac{1}{n}e^{-\frac{\pi}{E}n} = 
-\frac{\alpha}{\pi}E\sum_{n=1}^\infty\frac{1}{n}e^{-\frac{\pi}{E}n}\ .
\end{equation}
For small values we find approximately ($E \ll 1$):
\begin{equation}
\mathfrak{Im}\ \langle T_\mu^\mu\rangle(E)\approx -\frac{\alpha}{\pi}Ee^{-\frac{\pi}{E}}\ ,
\end{equation}
which goes to zero for $E\rightarrow 0$.

Our result can also be obtained by using the well-known formula (c.f. $e^+-e^-$ pair production)
\begin{equation}
\mathfrak{Im}\ \Leff^{(1)}(E) = \frac{\alpha}{2\pi^2}E^2\sum_{n=1}^\infty \frac{1}{n^2}e^{-\frac{\pi m^2}{eE}n}\quad .
\end{equation}
We only need to write
\begin{align*}
\mathfrak{Im}\ \langle T_\mu^\mu\rangle(E) &= m\frac{\partial}{\partial m}\mathfrak{Im}\ \Leff^{(1)}(E) 
= m\frac{\alpha}{2\pi^2}E^2\sum_{n=1}^\infty \frac{1}{n^2}\left(\frac{-2m\pi n}{eE}\right)e^{-\frac{\pi m^2}{eE}n}\\
&= -\frac{m^2 eE}{4\pi^2}\sum_{n=1}^\infty \frac{1}{n}e^{-\frac{\pi m^2}{eE}n}\quad ,
\intertext{or, in our units,}
\mathfrak{Im} \langle T_\mu^\mu\rangle(E) &= -\frac{\alpha}{\pi}E\sum_{n=1}^\infty \frac{1}{n}e^{-\frac{\pi}{E}n}\qquad \square
\end{align*}

Up to now, we have always restricted our calculations to the case of constant electric or magnetic fields. It can be shown, however, that the leading terms for strong fields,
i.e., those of order $H^2\ln H$ or $E^2\ln E$, are the same if the fields are not constant. This will be demonstrated in the next chapter, where we extend our discussion to the effective 
Yang-Mills field theory.

\section{The Effective Action for Yang-Mills Theory}\label{sec:EffActYangMills}
The effective action of quantum chromodynamics (QCD) for covariant constant color fields has been extensively treated in the literature by many researchers.
But they rarely pose the question in how far their results are physically reasonable and applicable. If one assumes that the confinement hypothesis is correct,
then no constant color fields can exist. Thus it would be physically senseless to study the effective Lagrangian (or the effective potential) in an
exact covariant constant color field. If, however, we regard a color field that in an expanded, but limited space, can be considered
to be approximately covariant constant, then one could suppose that in the space in question the effective Lagrangian can be approximated by the effective
Lagrangian of a covariant constant field. Thus we try to extrapolate from the case of an unlimited, expanded covariant color field to the case of a color field
that is in an expanded but limited space, approximately covariant constant. Upon looking more closely, it turns out that this procedure is physically unsatisfactory,
because one first calculates the effective Lagrangian for the covariant constant field configuration, which is not even theoretically feasible - this is forbidden by the
confinement hypothesis - and then tries to extrapolate to a physical situation. For such an extrapolation from a nonphysical to a physical situation, one cannot expect that the 
result is in any way physically acceptable. Thus the results obtained for the effective potentials with covariant constant color fields should not be used to describe
the nature of the QCD vacuum, but rather a transitional phase in the search for the true QCD vacuum.

After this prelude we will return to the role of the energy-momentum tensor in  Yang-Mills theory. Rather than attempt the difficult task of computing $\Leff^{eff}$ as 
done in \cite{Dittrich1983}, we will instead make an ansatz. Our ansatz will be motivated by the requirements that $\Leff^{eff}(x)$ gives the correct trace anomaly for the
energy-momentum tensor, and depends only on the algebraic invariant $F^2 := F^a_{\mu\nu}F^{a\mu\nu}$\cite{Pagels1978}.

So we require
\begin{align*}
\Theta^{\mu\nu} &= 2\frac{\Leff (F^2)}{\partial\eta_{\mu\nu}} - \eta^{\mu\nu}\Leff(F^2)\\
\Theta_\mu^\mu &= \frac{\beta(\bar{g}(t) )}{2\bar{g}^3(t)}F^a_{\mu\nu}F^{a\mu\nu}\ ,
\end{align*}
so that $\Theta_\mu^\mu$ has the usual form of the trace anomaly. Now we want to prove that these requirements are satisfied by the ansatz
\begin{equation}
\Leff^{eff} := -\frac{1}{4}\frac{1}{\bar{g}^2(t)}F^2\ ,\quad t:=\frac{1}{4}\ln\frac{F^2}{\mu^2}\quad .
\end{equation}
So we have to calculate
\begin{align*}
\Theta_{\mu\nu} &= 2\frac{\partial\Leff^{eff}}{\partial \eta^{\mu\nu}} - \eta_{\mu\nu}\Leff^{eff}\\
&= \left( 2\frac{1}{\bar{g}^2(t)}\frac{\partial}{\partial \eta^{\mu\nu}}\left(-\frac{1}{4}F^2\right) - \eta_{\mu\nu}\Leff\right) 
- \frac{2}{4}\left(\frac{\partial}{\partial\eta^{\mu\nu}}\frac{1}{\bar{g}^2(t)}\right)F^2\\
&= \frac{1}{\bar{g}^2(t)}\left[ \frac{1}{4}\eta_{\mu\nu}F^2-F_\mu^\alpha F_{\nu\alpha}\right] - \frac{1}{2}\left(\frac{\partial t}{\partial\eta^{\mu\nu}}\frac{d}{dt}\frac{1}{\bar{g}^2(t)}\right)F^2\\
&= \frac{1}{\bar{g}^2(t)}\left[ F_\mu^\alpha F_{\alpha\nu} + \frac{1}{4}\eta_{\mu\nu}F^2\right] + \frac{1}{4}\left(\frac{d}{dt}\frac{1}{\bar{g}^2(t)}\right)F^a_{\mu\beta}F^{a\beta}_\nu\ .
\end{align*}
The last term on the right-hand side uses the result
\begin{equation}
\frac{d}{dt}\frac{1}{\bar{g}^2(t)}= - \frac{2}{\bar{g}^3(t)}\frac{d\bar{g}}{dt}(t) = -2\frac{\beta(\bar{g}(t))}{\bar{g}^3(t)}\quad ,
\end{equation}
where we employed
\begin{equation}
t = \int_{g}^{\bar{g}(t)} \frac{dg'}{\beta(g')}\quad ,
\end{equation}
which, when taken the derivative $\tfrac{d}{dt}$ of, gives
\begin{equation}
1 = \frac{1}{\beta(\bar{g}(t))}\frac{d}{dt}\bar{g}(t),\text{ i.e., } \frac{d}{dt}\bar{g}(t) = \beta(\bar{g}(t))\quad .
\end{equation}
So we proved the relation
\begin{align*}
\Theta_{\mu\nu} &= \frac{1}{\bar{g}^2(t)}\left[F_\mu^\alpha F_{\alpha\nu} + \frac{1}{4}\eta_{\mu\nu}F^2\right] - \frac{\beta(\bar{g}(t))}{2\bar{g}^3(t)}F^a_{\mu\beta}F^{a\beta}_\nu\ ,
\intertext{and from here the trace}
\Theta_\mu^\mu &= \frac{\beta(\bar{g}(t))}{2\bar{g}^3(t)}F^a_{\mu\beta}F^{a\mu\beta}\ ,\quad t=\frac{1}{4}\ln\frac{F^2}{\mu^4}\quad .\quad \square
\end{align*}

We also can verify that our ansatz $\Leff^{eff} = -\frac{1}{4}\frac{F^2}{\bar{g}^2(t)},\ t=\frac{1}{4}\ln\frac{F^2}{\mu^4}$ satisfies the 
renormalization group equation
\begin{equation}
\left[\mu\frac{\partial}{\partial\mu}+\beta(g)\frac{\partial}{\partial g}\right]\left(-\frac{1}{4}F^2\right)\frac{1}{\bar{g}^2(t)} = 0\quad .
\end{equation}
Proof:
\begin{align*}
\mu\frac{\partial}{\partial\mu}\frac{1}{\bar{g}^2(t)} &= \mu\frac{\partial t}{\partial\mu}\frac{d}{dt}\frac{1}{\bar{g}^2(t)} 
= \mu\frac{1}{4}\frac{\partial}{\partial\mu}\left( \ln F^2 -4\ln\mu\right)(-2)\frac{\beta(\bar{g}(t))}{\bar{g}^3(t)}\\
&= 2 \frac{\beta(\bar{g}(t))}{\bar{g}^3(t)}\quad .
\end{align*}
Taking the derivative $\tfrac{\partial}{\partial g}$ of $t=\int_{g}^{\bar{g}(t)} \tfrac{dg'}{\beta(g')}$ we obtain
\begin{equation}
\frac{\partial}{\partial g}t = 0 = \frac{\partial\bar{g}(t)}{\partial g}\frac{1}{\beta(\bar{g}(t))}-\frac{1}{\beta(g)}\ \text{ or }\  \frac{\partial\bar{g}(t)}{\partial g} =
\frac{\beta(\bar{g}(t))}{\beta(g)}
\end{equation}
and from here
\begin{equation}
\beta(g)\frac{\partial}{\partial g }\frac{1}{\bar{g}^2(t)} = \beta(g)(-2)\bar{g}^{-3}(t)\frac{\partial\bar{g}(t)}{\partial g} 
= -2\frac{1}{\bar{g}^3(t)}\beta(g)\frac{\beta(\bar{g}(t))}{\beta(g)} = -2\frac{\beta(\bar{g}(t))}{\bar{g}^3(t)}\ .
\end{equation}
So we proved the equation
\begin{equation}
\left[\mu\frac{\partial}{\partial\mu}+\beta(g)\frac{\partial}{\partial g}\right]\frac{1}{\bar{g}^2(t)}=0\ .
\end{equation}

The effective Lagrangian is specified once we know $\beta(g)$. For weak coupling we have
\begin{align*}
g\beta(g) &= -\frac{1}{2}b_0g^4+b_1g^6+\dots\\
\beta(g) &= -\frac{1}{2}b_0g^3+b_1 g^5+\dots = -\frac{1}{2}b_0 g^3\left( 1- 2\frac{b_1}{b_0}g^2\right)+\dots \quad .
\end{align*}
This we substitute into
\begin{align*}
t &= \int_{g}^{\bar{g}(t)} \frac{dg'}{\beta(g')} = -\frac{2}{b_0}\int_g^{\bar{g}(t)} \frac{1}{g'^3(1-2\frac{b_1}{b_0}g'^2+\dots )}dg'\\
&= -\frac{2}{b_0}\int_{g}^{\bar{g}(t)}\left(\frac{1}{g'^3}+2\frac{b_1}{b_0}\frac{1}{g'}+\dots\right) dg'\\
&= -\frac{2}{b_0}\left[ -\frac{1}{2g'^2}+2\frac{b_1}{b_0}\ln g'+\dots\right]^{\bar{g}(t)}_{g}
\intertext{ to obtain}
t &= \frac{1}{b_0}\frac{1}{\bar{g}^2(t)}-\frac{2}{b_0}2\frac{b_1}{b_0}\ln\bar{g}(t)+\dots \underbrace{-\frac{1}{b_0}\frac{1}{g^2}+\frac{2}{b_0}2\frac{b_1}{b_0}\ln g+\dots}_{=\text{const.}\cdot t}\quad .
\end{align*}
Hence for $g(t)\ll 1$ we find the approximations
\begin{align*}
t &= \frac{1}{b_0}\frac{1}{\bar{g}^2_{(1)}(t)} &: \bar{g}_{(1)}(t) &= \sqrt{\frac{1}{b_0t}}\\
t &= \frac{1}{b_0}\frac{1}{\bar{g}^2_{(2)}(t)}\underbrace{-\frac{2}{b_0}2\frac{b_1}{b_0}\ln\sqrt{\frac{1}{b_0t}}}_{=\frac{1}{b_0}2\frac{b_1}{b_0}\ln t+\text{const.}}
&: \frac{1}{\bar{g}_{(2)}^2(t)} &=b_0t-2\frac{b_1}{b_0}\ln t\ .
\end{align*}
Consequently for large fields $F^2$, the effective Lagrangian is controlled by perturbation theory (asymptotic freedom) and is given by
\begin{equation}
\Leff^{eff}_{\text{leading log}} = -\frac{1}{16}b_0F^2\ln\frac{F^2}{\mu^4} + \mathcal{O}\left(F^2\ln\ln F^2\right)\quad .
\end{equation}
By the way, in QCD we have
\begin{align*}
\beta(g) &= \mu\frac{\partial g}{\partial\mu}\quad ,
\intertext{where}
\beta(g) &= -\frac{g^3}{16\pi^2}\left(\frac{11}{3}N-\frac{2}{3}N_f\right)+\dots\\
g\beta(g) &= -\frac{1}{2}g^4\frac{1}{8\pi^2}\left(\frac{11}{3}N-\frac{2}{3}N_f\right)+\mathcal{O}(g^6)\\
&= -\frac{1}{2}b_0g^4 + \mathcal{O}(g^6)\quad ,
\intertext{so that}
b_0 &= \frac{1}{8\pi^2}\left(\frac{11}{3}N-\frac{2}{3}N_f\right)\quad .
\end{align*}
Finally, let us write the leading-log effective Lagrangian in a form that will also be useful in QED:
\begin{align*}
\Leff^{eff} &= -\frac{1}{4}\frac{1}{\bar{g}^2(t)}F^2\ ,\\
t &:= \frac{1}{4}\ln\frac{F^2}{\mu^4}=\frac{1}{4}\ln\frac{2(B^2-E^2)}{\mu^4}\\
F^2 &:= F^a_{\mu\nu}F^{a\mu\nu} = -2\left(\vec{E}^a\cdot \vec{E}^a - \vec{B}^a\cdot\vec{B}^a\right) \equiv -2\left( E^2-B^2\right)\ .
\end{align*}
Then $\Leff^{eff}$ is given by
\begin{align*}
\Leff^{eff} &= -\frac{1}{4}b_0tF^2+\dots = -\frac{1}{4}b_0\frac{1}{4}\left[\ln\frac{F^2}{\mu^4}\right]F^2+\dots \\
 &= -\frac{1}{16}b_0F^2\ln\frac{F^2}{\mu^4}+\dots \\
 \Leff_{eff}^{(1)} &= \frac{1}{8}b_0(E^2-B^2)\ln\frac{2(B^2-E^2)}{\mu^4}+\dots\quad .
\end{align*}
When we rescale $g^2$ into the fields we obtain
\begin{equation}
\Leff^{eff} = \frac{1}{2}\left( \vec{E}^2-\vec{B}^2\right) + \frac{1}{2}\left( \vec{E}^2-\vec{B}^2\right)\frac{g^2}{48\pi^2}N\frac{11}{2}\ln\frac{g^2(\vec{B}^2-\vec{E}^2)}{\mu^4}
+\mathcal{O}(F^2)\ .
\end{equation}
This, by the way, is the same result that one obtains for covariant constant color fields.
Special cases are given by
\begin{align*}
\vec{E}=\vec{0}\quad :\quad \Leff^{eff}(B) &= -\frac{1}{2}B^2-\frac{1}{2}B^2\frac{g^2}{48\pi^2}N\frac{11}{2}\ln\frac{g^2B^2}{\mu^4}\\
&= -\frac{1}{2}B^2\left(1+\frac{g^2}{48\pi^2}11N\ln\frac{gB}{\mu^2}\right)\\
B\rightarrow\frac{1}{i}E\quad : \quad \Leff^{eff}(E) &= \frac{1}{2}E^2\left(1+\frac{g^2}{48\pi^2}11N\ln\frac{g(-iE)}{\mu^2}\right)\\
&= \frac{1}{2}E^2\left( 1+\frac{g^2}{48\pi^2}11N\left(\ln\frac{gE}{\mu^2} - \frac{i\pi}{2}\right)\right)\ .
\end{align*}
Since $V^{eff}(B) = -\Leff^{eff}(B)$ we obtain for the effective potential in QCD for $SU(N=3)$:
\begin{equation}
V^{eff}(B) = \frac{1}{2}B^2\left[ 1+\frac{g^2}{4\pi^2}\left(11\cdot 3 - 2N_f\right)\ln\frac{gB}{\mu^2}\right]\quad .
\end{equation}
Since $-\tfrac{1}{4}F^2=\tfrac{1}{2}(E^2-B^2)$, we have for the color magnetic fields only $B^2 = \tfrac{1}{2}F^2$:
\begin{align*}
V^{eff}(B) &= \frac{1}{4}F^2\left[ 1+\frac{g^2}{4}b_0\ln\frac{(gF)^2}{2\mu^4}\right]\\
b_0 &= \frac{1}{8\pi^2}\left(\frac{11}{3}\cdot 3 -\frac{2}{3}N_f\right)\  , \quad N=3\quad .
\end{align*}
To find the minimum we take the derivative $\tfrac{\partial V^{eff}(F)}{\partial F^2}$ and set it equal to zero. The result is
\begin{align*}
\ln\frac{e(gF)^2}{2\mu^4} &= -\frac{4}{b_0g^2}
\intertext{or}
\langle 0\vert (gF)^2\vert 0 \rangle &= \frac{2\mu^4}{e}e^{-\frac{4}{b_0g^2}}\quad ,
\intertext{which gives the dimensionless number}
\frac{\langle 0\vert (gF)^2\vert 0 \rangle}{\mu^4} &= \frac{2}{e}\underbrace{e^{-\frac{4}{b_0g^2}} }_{< 1}=0.7357e^{-\frac{4}{b_0g^2}}\quad .
\end{align*}
From here we find the expression for $V^{eff}_{min}(F)$ to be:
\begin{align*}
V^{eff}_{min}(F) &= \frac{1}{4}\langle F^2\rangle\left[1+\frac{g^2}{4}b_0\ln\frac{\langle (gF)^2\rangle}{2\mu^4}\right]\\
&= \frac{1}{4}\langle F^2\rangle \left[1+\frac{g^2}{4}b_0\left(\ln\frac{1}{e} + \ln e^{-\frac{4}{b_0g^2}}\right) \right]\\
&= -\frac{b_0}{16}\langle 0\vert (gF)^2\vert 0 \rangle \quad \left(=0.7124\cdot 10^{-2}\langle 0\vert (gF)^2\vert 0 \rangle \text{ for 3 massless flavours}\right)\\
&= -\frac{1}{128\pi^2}\left(11-\frac{2}{3}N_f\right)\langle 0\vert (gF)^2\vert 0 \rangle\\
&\underset{g^2=4\pi\alpha}{=} -\frac{1}{128\pi^2}\left(\frac{11}{3}N-\frac{2}{3}N_f\right)\langle 0\vert 4\pi\alpha F^2\vert 0 \rangle\\
&= -\frac{1}{32}\left(\frac{11}{3}N-\frac{2}{3}N_f\right)\langle 0 \vert \frac{\alpha}{\pi}F^a_{\mu\nu}F^{a\mu\nu}\vert 0\rangle + \mathcal{O}(\alpha^2)\quad .
\end{align*}
This result is consistent with the trace
\begin{equation}
\langle 0\vert T_\mu^\mu\vert 0 \rangle = \frac{1}{8}\left(\frac{11}{3}N-\frac{2}{3}N_f\right)\langle 0 \vert \frac{\alpha}{\pi}F^a_{\mu\nu}F^{a\mu\nu}\vert 0\rangle\ .
\end{equation}
From Lorentz invariance $T_{\mu\nu}(x)=\text{const.}\cdot g_{\mu\nu}$ and $T_{00} = \epsilon$; therefore $T_{\mu\nu}=\epsilon g_{\mu\nu}$, and we obtain
$T_\mu^\mu = 4\epsilon$ so that the lowering of the energy caused by non-perturbative fluctuations of the color field yields in the vacuum state
\begin{equation}
\epsilon = -\frac{1}{32}\left(\frac{11}{3}N-\frac{2}{3}N_f\right)\langle 0 \vert \frac{\alpha}{\pi}F^a_{\mu\nu}F^{a\mu\nu}\vert 0\rangle+\mathcal{O}(\alpha^2)\ .
\end{equation}
For $N=3,N_f=3$ we have $(\tfrac{11}{3}N-\tfrac{2}{3}N_f) = 9$.

\section{The Effective Lagrangian in QED}
We want to investigate the modification of Coulomb's law for long and short distances. First, we will ask for the effective Lagrangian
for weak, but otherwise arbitrary, fields.
In the weak-field limit, $e^2F_{\mu\nu}F^{\mu\nu}/m^4$ becomes small due to the smallness of $\alpha$. This leads to the expression
(wf = weak-field):
\begin{equation}
W_{wf}^{(1)}[A] = \int\Leff^{(1)}_{wf}d^4x = \frac{1}{2}\int A^\mu(x)\Pi_{\mu\nu}(x,y)A^\nu(y)d^4xd^4y
\end{equation}
for the weak-field limit of the one-loop effective action, where $\Pi_{\mu\nu}$ is nothing but the well-known order-$e^2$ polarization tensor of QED.

In momentum space it is given by
\begin{align*}
\Pi_{\mu\nu}(k) &= \left( g_{\mu\nu} k^2-k_\mu k_\nu\right)\Pi(k^2)\ ,\\
\Pi(k^2) &= -\frac{\alpha}{3\pi}k^2\int_{4m^2}^\infty \frac{1}{t}\rho(t)\frac{1}{k^2+t-i\epsilon}dt\ ,\\
\rho(t) &= \left(1+\frac{2m^2}{t}\right)\left(1-\frac{4m^2}{t}\right)^{\frac{1}{2}}\quad .
\end{align*}
As a consequence of the particular tensor structure of $\Pi_{\mu\nu}(k)$, $W_{wf}^{(1)}$ can be written in terms of $F_{\mu\nu}$ only and therefore is
gauge invariant.

Adding the classical Maxwellian term, we obtain for the real field effective Lagrangian
\begin{equation}
\Leff^{eff}_{wf} = -\frac{1}{4}F_{\mu\nu}(x)\left[1+\frac{\alpha}{3\pi}\square\int_{4m^2}^\infty \frac{1}{t}\frac{\rho(t)}{t-\square}dt\right]F^{\mu\nu}(x)
\end{equation}
where, as usual, $\square = -\partial_t^2+\vec{\nabla}^2$.

The equations of motion resulting from this effective Lagrangian for the weak-field case are linear, because $\Leff^{eff}_{wf}$ is quadratic in the fields.
Next, let us apply $\Leff^{eff}_{wf}$ to the Coulomb problem.
Specializing to the static case begins with the variation
\begin{equation}
\frac{\delta}{\delta A^0(\vec{x})}\int d^3x'\left[ \frac{1}{2} \vec{\nabla} A^0\cdot \left( 1+\frac{\alpha}{3\pi}\vec{\nabla}^2\int_{4m^2}^\infty \frac{1}{t}\frac{\rho(t)}{t-\vec{\nabla}^2}dt\right)
\vec{\nabla} A^0-A^0J_0\right] = 0\ .
\end{equation}
For $J_0(\vec{x})$ we assume two point charges with the separation $r$:
\begin{equation}
J_0(\vec{x})=Q\left[\delta\left(\vec{x}-\vec{x}_1\right) - \delta\left(\vec{x}-\vec{x}_2\right)\right]\ ,\quad |\vec{x}_1-\vec{x}_2| = r\ .
\end{equation}
The variation then gives the equation of motion
\begin{align*}
D\vec{\nabla}^2 A^0(\vec{x}) &= -J_0(\vec{x})\quad ,
\intertext{where}
D &:= 1+\frac{\alpha}{3\pi}\vec{\nabla}^2\int_{4m^2}^\infty \frac{1}{t}\frac{\rho(t)}{t-\vec{\nabla}^2}dt\quad .
\end{align*}
Making use of the position space representation of the resolvent $(t-\vec{\nabla}^2)^{-1}$, one can easily calculate the potential energy 
$V=-\int\Leff_{wf}^{eff}(A^0)d^3\vec{x}$ associated with the interaction of two point charges. One finds
\begin{equation}
V(r) = -\frac{Q^2}{4\pi}\left[\frac{1}{r}+\frac{\alpha}{3\pi}\int_{4m^2}^\infty \frac{\rho(t)}{t}\frac{e^{-\sqrt{t}r}}{r}dt\right]+\mathcal{O}(\alpha^2)\ .
\end{equation}
The second term in the brackets is the well-known Uehling correction to the classical Coulomb potential. $V(r)$ was derived in the weak-field limit and thus should be
valid at large distances. Because the equation of motion is linear, $V(r)$ takes the form of a superposition of Yukawa potentials.

The quantum mechanical correction to a many-particle static potential,
\begin{align*}
A^0(\vec{x}) &= \sum_i \frac{Q_i}{4\pi|\vec{x}-\vec{x}_i|}\ ,\quad J_0(\vec{x}) = \sum_iQ_i\delta(\vec{x}-\vec{x}_i)\\
V_{static} &= \frac{1}{2}\sum_{i\neq j}\frac{Q_iQ_j}{4\pi|\vec{x}_i-\vec{x}_j|}\ , \quad r_{ij} = |\vec{x}_i-\vec{x}_j|\ ,\\
\intertext{becomes}
V_{static} &= \frac{1}{2}\sum_{i\neq j}\frac{Q_iQ_j}{4\pi}\left[\frac{1}{r_{ij}}+\frac{\alpha}{3\pi}\int_{4m^2}^\infty \frac{\rho(t)}{t}\frac{e^{-\sqrt{t}r_{ij}}}{r_{ij}}dt
\right]+\mathcal{O}(\alpha^2)\ .
\end{align*}
Now that we have established the Lagrangian $\Leff^{eff}_{wf}$ and the (one-loop) correction to the Coulomb potential for weak fields, we want to set up the 
generalized Maxwell equations for strong fields. The following calculations are justified by noting that the ansatz
\begin{equation}
\Leff_{eff} = -\frac{1}{4e^2(F^2)}F_{\mu\nu}F^{\mu\nu}
\end{equation}
leads to the correct trace anomaly of the energy-momentum tensor (c.f. \ref{sec:EffActYangMills})

One starts from the free Maxwell Lagrangian 
\begin{equation}
\Leff = -\frac{1}{4}F_{\mu\nu}(x)F^{\mu\nu}(x)
\end{equation}
and scales the electromagnetic coupling $e$ out of the fields
\begin{align*}
A_\mu &\rightarrow \frac{1}{e}A_\mu\quad ,
\intertext{giving}
\Leff &= -\frac{1}{4e^2}F_{\mu\nu}F^{\mu\nu}\ .
\end{align*}
Note that in the complete interacting QED Lagrangian this is the only term containing $e$, because the vertex now simply
reads $\bar{\psi}\gamma\cdot A\psi$ instead of $e\bar{\psi}\gamma\cdot A\psi$.

The next step is to renormalization-group improve $\Leff$ by replacing $e$ with the running coupling constant $e(\mu)$ to first
order in $\alpha$.
To achieve this we make use of the scale variation of the gauge coupling constant\cite{Ramond1997}
\begin{equation}
\mu\frac{\partial e}{\partial\mu} = \beta(e) = \frac{1}{12\pi^2}e^3\ .
\end{equation}
This equation is solved by
\begin{equation}
\frac{1}{e^2(\mu)}-\frac{1}{e^2(\mu_0)} = -\frac{1}{6\pi^2}\ln\frac{\mu}{\mu_0}\ ,
\end{equation}
where $\mu_0$ denotes an arbitrary scale.

To prove this let us rewrite the last equation in the form
\begin{align*}
\frac{1}{e^2(\mu)} &= \frac{1}{e^2(\mu_0)} -\frac{1}{6\pi^2}\ln\frac{\mu}{\mu_0}\\
\text{or}\qquad e(\mu) &= \left(\frac{1}{e^2(\mu_0)} -\frac{1}{6\pi^2}\ln\frac{\mu}{\mu_0}\right)^{-\frac{1}{2}}\ .
\intertext{Hence}
\frac{\partial e(\mu)}{\partial \mu} &= -\frac{1}{2} \left(\frac{1}{e^2(\mu_0)} -\frac{1}{6\pi^2}\ln\frac{\mu}{\mu_0}\right)^{-\frac{3}{2}}
\left(-\frac{1}{6\pi^2}\right)\frac{1}{\mu}\ ,
\intertext{which yields}
\mu\frac{\partial e(\mu)}{\partial\mu} &= \frac{1}{12}e^3\qquad \square
\end{align*}
Let us rewrite our solution slightly:
\begin{align*}
\frac{1}{e^2(\mu_0)}\left(\frac{e^2(\mu_0)}{e^2(\mu)}-1\right) &= -\frac{1}{6\pi^2}\ln\frac{\mu}{\mu_0}\\
\text{or}\qquad\frac{e^2(\mu_0)}{e^2(\mu)} &= 1-\frac{e^2(\mu_0)}{6\pi^2}\ln\frac{\mu}{\mu_0}\ .
\end{align*}
So the scaling equation for $e^2(\mu) := 4\pi\alpha(\mu)$ is given by
\begin{equation}
e^2(\mu) = \frac{e^2(\mu_0)}{1-\frac{e^2(\mu_0)}{6\pi^2}\ln\frac{\mu}{\mu_0}}\ .
\end{equation}
This equation has a singularity which follows from
\begin{align*}
1-\frac{e^2(\mu_0)}{6\pi^2}\ln\frac{\mu}{\mu_0} &= 0\\
\text{or}\qquad \frac{e^2(\mu_0)}{6\pi^2}\ln\frac{\mu}{\mu_0} &= 1\ ,
\end{align*}
which is solved by $\mu = \mu_0 e^{\frac{6\pi^2}{e^2(\mu_0)} }$ and is known as Landau singularity.
Our scaling equation underlines the fact that the electric charge grows weaker and weaker at large distances (i.e., small scales),
which means that the identification of the free Lagrangian ($e=0$) in terms of physical photons is perfectly justified.

In our application, where the fields are sufficiently strong so that fermionic masses are negligible, the length or mass scale is set by the
magnitude $F^2 = F_{\mu\nu}F^{\mu\nu}$. Therefore we replace in the scaling equation $\mu^4$ by $F^2$ to obtain
\begin{equation}
e^2(F^2) = \frac{e^2(\mu_0)}{1-\frac{e^2(\mu_0)}{24\pi^2}\ln\frac{F^2}{\mu_0^4}}\ ,
\end{equation}
with an arbitrary integration constant and arbitrary reference mass $\mu_0$. After replacing $e^2$ by the field-dependent running coupling
constant $e^2(F^2)$ we obtain
\begin{align*}
\Leff_{eff} &= -\frac{1}{4e^2(F^2)}F_{\mu\nu}F^{\mu\nu}\\
&= -\frac{1}{4e^2(\mu_0)}F_{\mu\nu}F^{\mu\nu}\left[ 1- \frac{e^2(\mu_0)}{24\pi^2}\ln\frac{F^2}{\mu_0^4}\right]\ .
\end{align*}
The one-loop part is (we scale back $e^2(\mu_0) := e^2$ into the fields)
\begin{align*}
\Leff^{(1)} &= \frac{1}{4}F_{\mu\nu}F^{\mu\nu}\frac{e^2}{24\pi^2}\ln\frac{e^2F^2}{\mu_0^4}\\
&\overset{F^2=-2(\vec{E}^2-\vec{B}^2)}{=} \frac{1}{2}\left(\vec{B}^2-\vec{E}^2\right)\frac{e^2}{24\pi^2}\ln\frac{e^2\left(\vec{B}^2-\vec{E}^2\right)}{\mu_0^4}
+\mathcal{O}(F^2)\\
\vec{E}=\vec{0}:\qquad \Leff^{(1)}(B) &= \frac{1}{2}B^2\frac{e^2}{24\pi^2}\ln\frac{(eB)^2}{\mu_0^4} = \frac{\alpha B^2}{6\pi}\ln\frac{eB}{\mu_0^2}\\
B\rightarrow\frac{1}{i}E:\qquad \Leff^{(1)}(E) &= -\frac{\alpha E^2}{6\pi}\ln\frac{e(-iE)}{\mu_0^2}\\
&= -\frac{\alpha E^2}{6\pi}\ln\frac{eE\cdot e^{-i\frac{\pi}{2}} }{\mu_0^2} = -\frac{\alpha E^2}{6\pi}\left[\ln\frac{eE}{\mu_0^2}-\frac{\pi}{2}i\right]\\
\mu_0 \equiv m:\qquad\Leff^{(1)}(E) &= -\frac{\alpha E^2}{6\pi} \left(\ln\frac{eE}{m^2}-\frac{\pi}{2}i\right)\quad ,
\end{align*}
and this is our old formula for $\Leff^{(1)}\left(\frac{eE}{m^2}\rightarrow\infty\right)$ which was formally derived for constant fields only.
However, at no point in the above \textquotedblleft derivation\textquotedblright\ of $\Leff^{(1)}$ did we have to demand the fields to be
constant; thus we may assume that $\Leff^{(1)}$ is correct to order $F^2\ln F$ for arbitrary varying fields.

Now that we have established the Lagrangian for strong but otherwise arbitrary fields, we can set up the generalized Maxwell equations and try to solve them
for a given source distribution. In general, they are of the form
\begin{equation}
\frac{\delta}{\delta A_\mu(x)}\int\left[\Leff^{(0)}+\Leff^{(1)}-J_\mu A^\mu \right]d^4x' = 0\ ,
\end{equation}
with $\Leff^{(1)}$ given by its real part. Furthermore $J_\mu (x) = J_0(\vec{x})\delta_\mu$ and $\vec{E}(\vec{x}) = -\vec{\nabla} A^0(\vec{x})\ (A^0=\phi)$.
This leads us to evaluating
\begin{equation}
\frac{\delta}{\delta A^0(\vec{x})}\int \left(\frac{1}{2}|\vec{\nabla} A^0|^2\left[1-\frac{e^2}{12\pi^2}\ln\frac{e|\vec{\nabla} A^0|}{m^2}\right]-J^0A^0\right) d^3\vec{x}' = 0\ ,
\end{equation}
which is equivalent to
\begin{equation}
V_{static} = -\text{ext.}_{\phi} \int\left( \frac{1}{2}(\vec{\nabla} \phi)^2\left[ 1 -\frac{\alpha}{3\pi}\ln\frac{e|\vec{\nabla}\phi|}{m^2}\right] -\phi J_0\right)d^3x\ .
\end{equation}
The variation of our $\phi$ then gives the local nonlinear differential equation for $\phi \equiv A^0$:
\begin{align*}
\partial_k \frac{\partial \Leff}{\partial(\partial_k\phi)} &= \frac{\partial \Leff}{\partial \phi}\quad ,
\intertext{where}
\Leff(\phi,\partial_i\phi) &= \frac{1}{2}\partial_k\phi\partial_k\phi\left[1-\frac{\alpha}{6\pi}\ln\frac{e^2(\partial_k\phi)(\partial_k\phi)}{m^4}\right]-\phi J_0\\
\frac{\partial\Leff}{\partial\phi} &= -J_0\\
\frac{\partial\Leff}{\partial(\partial_k\phi)} &= \underbrace{\partial_k\phi}_{=-E_k}\left[1-\frac{\alpha}{6\pi}\ln\frac{e^2(\partial_k\phi)(\partial_k\phi)}{m^4}\right]
+\left( -\frac{\alpha}{6\pi}\partial_k\phi\right)\ .
\intertext{Therefore}
\partial_k\frac{\partial\Leff}{\partial(\partial_k\phi)} &= \partial_k\left((-E_k)\left[1-\frac{\alpha}{3\pi}\ln\frac{eE}{m^2}\right]+\left( \frac{\alpha}{6\pi}E_k\right)\right)
=-J_0\\
\text{and thus}\ &\vec{\nabla}\cdot\left[\left(1-\underbrace{\frac{\alpha}{6\pi}}_{\ll \frac{\alpha}{3\pi}\ln\frac{eE}{m^2}} - \frac{\alpha}{3\pi}\ln\frac{eE}{m^2}\right)\vec{E}\right] = J_0\ .
\end{align*}
Alltogether we have
\begin{align*}
\vec{\nabla}\cdot \vec{D} &= J_0  ,& \vec{D} &:=\epsilon(E)\vec{E}\\
\epsilon(E) &= 1-\frac{\alpha}{3\pi}\ln\frac{eE}{m^2} & \vec{E} &= -\vec{\nabla}\phi
\end{align*}
These are well-known classical equations from electrostatics of polarizable media. Looking back at the microscopic origin of $\epsilon(E)$,
we see that the effect of the vacuum fluctuations of the electron field is such that the vacuum responds to an external electric field as
if it were some sort of crystal which possesses a field-dependent dielectric constant. Obviously, Maxwell's equations become nonlinear due to the
logarithm in $\epsilon(E)$.

To summarize, we can say that in deriving $\vec{\nabla}\cdot \vec{D} = J_0$ with $\vec{D}=\epsilon(E)\vec{E}$, we have solved the problem of finding the nonlinear generalization of Maxwell's
equations - the non-linearities being caused by the electrons, which are hidden from direct observation but which influence the dynamics of the $A_\mu$ field
for a strong and static, but otherwise arbitrary, electrical field.

To gain some insight into the effect produced by the nonlinear medium $\epsilon(r)$, let us look at a specific example. We consider the case where $J_0$
contains only a single isolated charge $Q$ at $\vec{x}=\vec{0}$ (together with a compensating spherical shell of charge $-Q$ at infinity):
\begin{equation}
J_0(\vec{x}) = Q\delta(\vec{x})\ .
\end{equation}
Making the spherically symmetric ansatz
\begin{equation}
\vec{D}=\frac{Q}{4\pi r^2}\hat{r},\quad \vec{E} = \frac{Q(r)}{4\pi r^2}\hat{r},\quad r = |\vec{r}|,
\end{equation}
the equation
\begin{equation}
\vec{\nabla}\cdot \vec{D} = Q\delta(\vec{x})
\end{equation}
is solved, provided that the function $Q(r)$ is a solution to the transcendental equation
\begin{equation}
Q = Q(r)\epsilon\left(\frac{eQ(r)}{4\pi r^2}\right)\ .
\end{equation}
The physical interpretation of $Q(r)$ is that it is the charge lying within a sphere of radius $r$ centered at $\vec{x}=\vec{0}$. The value of $Q(r)$
is always larger than $Q$ because the vacuum polarization effects screen the charge. If we let $r\rightarrow\infty,\ Q(r)$ approaches the (macroscopically)
observed charge $Q$. We thus got an implicit equation for the modification of Coulomb's law by the electron fluctuations:
\begin{equation}
E(r) = \frac{Q(r)}{4\pi r^2}\ .
\end{equation}
One can show that when
\begin{equation}
\frac{eQ(r)}{4\pi}\gg 1\quad ,
\end{equation}
the approximation of neglecting the nonlocal one-loop contribution in the effective action becomes self-consistent. To see why this should be so, we note
that because of the leading local but nonlinear correction to $\epsilon$,
\begin{equation}
-\frac{\alpha}{3\pi}\ln X_1\ ,\quad X_1 = \frac{eQ(r)}{4\pi r^2m^2}\ ,
\end{equation}
while for the leading nonlocal correction
\begin{equation}
-\frac{\alpha}{3\pi}\ln X_2\ ,\quad X_2 = \frac{\left(\vec{\nabla}\ln A^0\right)^2}{m^2}\sim\frac{1}{r^2m^2}\ ,
\end{equation}
so that $\tfrac{eQ(r)}{4\pi}\gg 1$ is just the condition for $X_1 \gg X_2$, i.e., nonlinear local effects win out at short distances.

\section{Adler's Leading-Log Model in QCD}
\begin{figure}
\centering
\includegraphics[width=0.6\linewidth]{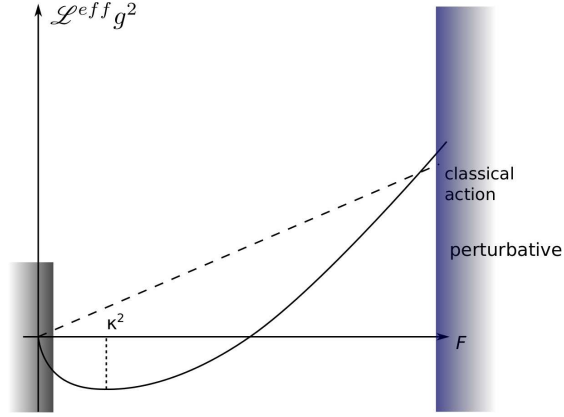}
\caption{Schematic graph of $\Leff^{eff}$. The dashed line represents classical action.}
\label{fig:EffLagrangian}
\end{figure}

Here, the leading approximation to the effective action is obtained by replacing $g^2$ in the classical Lagrangian
\begin{equation}
\Leff = \frac{1}{2g^2}\left(\vec{E}^2-\vec{B}^2\right)\ ,
\end{equation}
by the running coupling constant
\begin{align*}
\Leff^{eff} &= \frac{1}{2g^2_{running}}\left(\vec{E}^2-\vec{B}^2\right)\ ,
\intertext{where}
g^2_{running}\left(\frac{\vec{E}^2-\vec{B^2}}{\mu^4}\right) &= \frac{g^2(\mu^2)}{1+\frac{1}{4}b_0g^2(\mu^2)\ln\left(\frac{\vec{E}^2-\vec{B^2}}{\mu^4}\right)}\ ,
\intertext{so that}
\Leff^{eff} &= \frac{1}{2g^2}\left(\vec{E}^2-\vec{B}^2\right)\left[1+\frac{1}{4}b_0g^2(\mu^2)\ln\left(\frac{\vec{E}^2-\vec{B^2}}{\mu^4}\right)\right]\ .
\end{align*}
Here $\mu$ is an arbitrary subtraction point, $g^2=g^2(\mu^2)$, and $b_0$ is a certain constant one gets from calculating the one-loop radiative corrections,
\begin{equation}
b_0 = \frac{1}{8\pi^2}\frac{11}{3}C_{ad} > 0\ .
\end{equation}
Our formula for $\Leff^{eff}$ is applicable if $g^2_{running}$ is small:
\begin{align*}
\left(\vec{E}^2-\vec{B}^2\right) &\gg \mu^4\\
\vec{E},\vec{B} &\approx \vec{0}\ :\ g^2\text{ small, }g^2<0\ .
\end{align*}
The variational equation will be
\begin{equation}
\frac{\delta W^{eff}}{\delta(E,B)} = 0\quad ,
\end{equation}
and the static potential follows from 
\begin{equation}
V_{static} = -W^{eff}_{extremum} + \text{self energies}\ .
\end{equation}
A reminder:
\begin{align*}
\Leff &= -\frac{1}{4}F_{\mu\nu}F^{\mu\nu}+j_\mu A^\mu\\
\vec{j}=\vec{0}\quad W &= \int\left[ \frac{1}{2}\left(\vec{E}^2-\vec{B}^2\right) - \rho\varphi\right] d^3x,\quad W=\frac{\text{action}}{T}\\
\frac{\delta W}{\delta B}=0\Rightarrow \vec{B}=\vec{0}\quad W &= \int\left[ \frac{1}{2}(\vec{\nabla}\varphi)^2-\rho\varphi\right] d^3x\\
\delta W &= \int\left[ \vec{\nabla}\varphi\cdot \vec{\nabla}\delta\varphi - \rho\delta\varphi\right] d^3x\\
&= \int \left(-\vec{\nabla}^2\varphi - \rho\right)\delta\varphi d^3x = 0\quad .
\end{align*}
Thus $\varphi_{extr.}$ satisfies $\vec{\nabla}^2\varphi_{extr.} = -\rho$.
\begin{align*}
W_{extr.} := W[\varphi_{extr.}]  &= \int \left[\frac{1}{2}\vec{\nabla}\varphi\cdot \vec{\nabla}\varphi-\rho\varphi\right]d^3x\\
&= \int \left[-\frac{1}{2}\varphi\vec{\nabla}^2\varphi-\rho\varphi\right]d^3x\\
&= \int \left[-\frac{1}{2}\varphi(-\rho)-\rho\varphi\right]d^3x\\
&= -\frac{1}{2}\int \rho\varphi d^3x = -V_{static}+\text{self energies}.
\end{align*}
Before we continue, let us rewrite $\Leff^{eff}$ in a more compact form (here we follow Adler and use $F$ instead of $F^2$ so far).
Using $F:=\vec{E}^2-\vec{B}^2,\quad \vec{E} = -\vec{\nabla}\varphi,\quad \vec{B}=\vec{\nabla}\times\vec{A}$ we have
\begin{align*}
\Leff^{eff}(F) &= \frac{1}{8}b_0F\left[\frac{4}{g^2(\mu^2)b_0}+\ln\frac{F}{\mu^4}\right]\\
&= \frac{1}{8}b_0F\left[ -\ln e^{-\frac{4}{g^2(\mu^2)b_0} } +\ln\frac{F}{\mu^4}  \right]\\
&= \frac{1}{8}b_0F\ln\frac{F}{\mu^4e^{-4/g^2(\mu^2)b_0} }\\
\Leff^{eff} &= \frac{1}{8}b_0\ln\frac{F}{e\kappa^2}\ ,
\intertext{where $\kappa^2$ is the constant}
\kappa^2 &:= \frac{\mu^4}{e}e^{-\frac{4}{b_0g^2(\mu^2)} }\ .
\end{align*}
This constant, $\kappa$, is a combination of $\mu$ and of $g^2(\mu^2)$. However, $\kappa^2$ is renormalization-group invariant (to one-loop order), so that 
$\kappa$ is a physical parameter, whereas $\mu$ is an unphysical parameter. We recall:
\begin{equation}
\beta(g) = \mu\frac{\partial g}{\partial\mu} = -\frac{1}{2}g^3b_0\quad .
\end{equation}
Then we obtain
\begin{align*}
\frac{d}{d\mu}\kappa^2 &= \frac{1}{e}\left[ 4\mu^3e^{-\frac{4}{b_0g^2}}+\mu^4\frac{d}{d\mu}\left(-\frac{4}{b_0g^2}\right)e^{-\frac{4}{b_0g^2}}\right]\\
&= \frac{1}{e}e^{-\frac{4}{b_0g^2}}\left[4\mu^3-\frac{4\mu^4}{b_0}(-2)g^{-3}\frac{d g}{d\mu}\right]\\
&= \frac{1}{e}e^{-\frac{4}{b_0g^2}}4\mu^3\left[1+\frac{2}{b_0g^3} \underbrace{\mu\frac{\partial g}{\partial\mu}}_{-\frac{b_0}{2}g^3} \right] = 0\quad .
\end{align*}
The graph of $\Leff^{eff}$ is shown in fig. \ref{fig:EffLagrangian}.

The minimum of $\Leff^{eff}$ is given by
\begin{align*}
\left( F\ln\frac{F}{e\kappa^2}\right)' &= \left( F\ln F - F\ln e\kappa^2\right)' = \ln F+1-\ln e\kappa^2\\
&= \ln F -\ln\kappa^2 = 0\\
\Rightarrow F_{min} &= \kappa^2 = \frac{\mu^4}{e}e^{-\frac{4}{b_0g^2(\mu^2)} },\quad b_0>0\quad .
\end{align*}
The renormalization group argument says that $\Leff^{eff}(F)$ is a good approximation in a region for strong fields and a region close to the origin for very weak fields.
Around the minimum, the approximation is not reliable. But what is reliable is the fact that the minimum is away from the origin where the interesting structure of the model comes from.
Now we have
\begin{equation}
W^{eff} = \int\left[ \Leff^{eff} - \rho\varphi\right] d^3x\ ,
\end{equation}
with
\begin{equation}
\rho(\vec{x}) = Q\left[\delta^{(3)}(\vec{x}-\vec{x}_1) - \delta^{(3)}(\vec{x}-\vec{x}_2)\right]\ .
\end{equation}
The variational equations come from
\begin{align*}
& W^{eff} &= \int\left[ \Leff^{eff}(\vec{\nabla}\varphi, \vec{A})-\rho\varphi\right] d^3x & \\
\Leff \equiv \Leff^{eff}:\quad \vec{\nabla}\cdot \frac{\partial\Leff}{\partial(\vec{\nabla}\varphi)}-\frac{\partial\Leff}{\partial\varphi} &= 0 & 
\frac{1}{2} F = \frac{1}{2}\left(\vec{E}^2-\vec{B}^2\right) &= \frac{1}{2}\left[(\vec{\nabla}\varphi)^2-\left(\vec{\nabla}\times\vec{A}\right)^2\right]\\
\text{or}\quad\vec{\nabla}\cdot\underbrace{\frac{\partial\Leff}{\partial\frac{1}{2}F}}_{=:\epsilon}\underbrace{\frac{\partial \frac{1}{2}F}{\partial(\vec{\nabla}\varphi)}}_{=-\vec{E}}+\rho &= 0 &
\frac{\partial\frac{1}{2}F}{\partial(\vec{\nabla}\varphi)} = \vec{\nabla}\varphi &= -\vec{E}\\
\Longrightarrow \vec{\nabla}\cdot\epsilon\vec{E} &= \rho & &
\end{align*}
\begin{align*}
\frac{\delta\Leff(x)}{\delta A_i(z)} &= \int \frac{\partial\Leff(x)}{\partial B_j(x)}\frac{\delta B_j(x)}{\delta A_i(z)}dx\\
&= \int \frac{\partial\Leff(x)}{\partial\frac{1}{2}F(x)}\frac{\partial\frac{1}{2}F}{\partial B_j(x)}\frac{\delta}{\delta A_i(z)}\epsilon_{jmn}\partial_m A_n(x)dx\\
&= \int (-\epsilon B_j)(x)\epsilon_{jmn}\delta_{in}\partial_m\delta(x-z)dx\\
&= \int \partial_m(\epsilon B_j)(x)\epsilon_{jmi}\delta(x-z)dx\\
&= \epsilon_{ijm}\partial_m(\epsilon B_j)(z) = -\left[\vec{\nabla}\times(\epsilon\vec{B})\right]_i = 0\\
\Rightarrow\quad \vec{\nabla}\times(\epsilon\vec{B}) &= \vec{0}
\end{align*}
\begin{align}
\epsilon(F) := \frac{\partial\Leff^{eff}}{\partial(\frac{1}{2}F)} &= \frac{1}{4}b_0\left(F\ln\frac{F}{e\kappa^2}\right)'\notag\\
\Rightarrow\quad \epsilon(F) &= \frac{1}{4}b_0\ln\frac{F}{\kappa^2}\ .\label{eq:PermettivityFunction}
\end{align}
\begin{figure}
\centering
\includegraphics[width=0.5\linewidth]{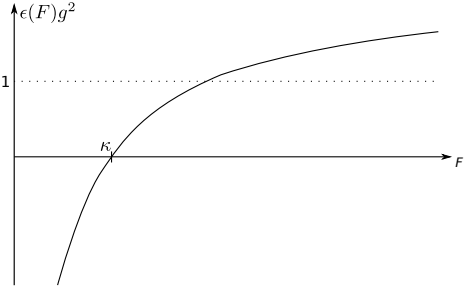}
\caption{Scaled permettivity function $\epsilon$ as given in Eq. \eqref{eq:PermettivityFunction}.}
\end{figure}
Now the equation
\begin{equation}
\vec{\nabla}\times(\epsilon\vec{B}) = \vec{0}
\end{equation}
is solved by $\epsilon\vec{B}=\vec{0}$.

Thus there are two branches we have to consider here:
\begin{align}
\vec{B} &= \vec{0},\quad \text{ or }\\
\epsilon &= 0,\quad \text{i.e., at }F=\vec{E}^2-\vec{B}^2=\kappa^2\ .
\end{align}
Near the source charges, where the fields are strong, asymptotic freedom tells us that the solution will look like the Abelian case. This means the electric field be big
and the magnetic field should be zero, or small:
\begin{align*}
\vec{B}=\vec{0}\ :\quad F=\vec{E}^2 &= (\vec{\nabla}\varphi)^2\ ,\ \vec{E}=-\vec{\nabla}\varphi\ .
\intertext{Now define}
\vec{D} &=\epsilon(\vec{E}^2)\vec{E} = \epsilon((\vec{\nabla}\varphi)^2)\vec{E}\quad ,
\intertext{together with}
\vec{\nabla}\cdot\vec{D} &= \rho\\
\vec{\nabla}\times\vec{E} &= \vec{0}\ .
\end{align*}
Thus, we now have a non-linear dielectric problem.

As S. Adler\cite{Adler1981} has shown, the leading-log model gives us a qualitatively correct, and semiquantitatively accurate account of the $q\bar{q}$ force. He shows in a highly non-trivial 
calculation that as the distance between the quarks $R=|\vec{x}_1-\vec{x}_2|\rightarrow 0,\infty$, one obtains
\begin{align*}
V_{static} (R) &\rightarrow \kappa QR + \mathcal{O}\left(\sqrt{\kappa}\ln(\sqrt{\kappa}R)\right),\quad R\rightarrow\infty\\
V_{static}(R) &\rightarrow -\frac{Q^2}{4\pi R\frac{1}{2}b_0}\left[\frac{1}{\ln\left(\frac{1}{\Lambda_p^2R^2}\right)}+\mathcal{O}\left(\frac{\log\log}{\log^2},\frac{1}{\log^3}\right)\right],\ 
R\rightarrow 0\ ,
\end{align*}
with $\Lambda_p = 2.52\sqrt{\kappa}$ for the parameter values $Q=\sqrt{\frac{4}{3}},b_0 = \frac{9}{8\pi^2}$ appropriate to $SU(3)$ with $3$ light quark flavors.

\bibliography{ArticleReferences.bbl}

\begin{thebibliography}{Ram97}

\bibitem[Adl81]{Adler1981}
Stephen~L. Adler.
\newblock Effective-action approach to mean-field non-abelian statics, and a
  model for bag formation.
\newblock {\em Physical Review D: Particles and Fields}, 24, 1981.

\bibitem[DR83]{Dittrich1983}
W.~Dittrich and M.~Reuter.
\newblock Effective {QCD}-lagrangian with $\zeta$-function regularization.
\newblock {\em Physics Letters B}, 128(5):321--326, September 1983.

\bibitem[DR85]{Dittrich1985}
Walter Dittrich and Martin Reuter.
\newblock {\em Effective Lagrangians in Quantum Electrodynamics}.
\newblock Lecture Notes in Physics. Springer, 1985.

\bibitem[PT78]{Pagels1978}
H.~Pagels and E.~Tomboulis.
\newblock Vacuum of the quantum {Y}ang-{M}ills theory and magnetostatics.
\newblock {\em Nuclear Physics B}, 143(3):485--502, October 1978.

\bibitem[Ram97]{Ramond1997}
Pierre Ramond.
\newblock {\em Field Theory: A Modern Primer}.
\newblock Westview Press, second edition, 1997.

\bibitem[Sch51]{Schwinger1951}
Julian Schwinger.
\newblock On gauge invariance and vacuum polarization.
\newblock {\em Physical Review}, 82(5):664--679, June 1951.

\end{thebibliography}
\end{document}